# Unveiling a crystal's entropy of disorder via electron diffraction: A statistical mechanics approach

Dongxin Liu[1], Oren Elishav[2], Jiarui Fu[1], Masaya Sakakibara[1], Kaoru Yamanouchi[1], Barak Hirshberg[2,3]*, Takayuki Nakamuro[1]*, Eiichi Nakamura[1]*

[1]Department of Chemistry, The University of Tokyo; 7-3-1 Hongo, Bunkyo-ku, Tokyo 113-0033, Japan

[2]School of Chemistry, Tel Aviv University, Tel Aviv 6997801, Israel

[3]The Center for Computational Molecular and Materials Science, Tel Aviv University, Tel Aviv 6997801, Israel.

*Corresponding author. hirshb@tauex.tau.ac.il (BH), muro@chem.s.u-tokyo.ac.jp (TN), nakamura@chem.s.u-tokyo.ac.jp (EN)

## Abstract

Upon melting, the molecules in the crystal explore numerous configurations, reflecting an increase in disorder. The molar entropy of disorder can be defined by Bolzmann's formula $\Delta S_d = R\ln(W_d)$ where $W_d$ is the increase in the number of microscopic states, so far inaccessible experimentally. We found that the Arrhenius frequency factor $A$ of the electron diffraction signal decay provides $W_d$ via an experimental equation $A = A_{INT}W_d$ where $A_{INT}$ is an inelastic scattering cross-section. The method connects Clausius and Boltzmann experimentally and supplements the Clausius approach, being applicable to a femtogram quantity of thermally unstable and biomolecular crystals. The data also showed that crystal disordering and crystallization of melt are reciprocal, both governed by the entropy change, but manifesting in opposite directions.

**Main Text**

When a molecular crystal melts, the molecules explore a larger number of configurations, reflecting an increase in disorder. Entropy change is a measure of disordering in a system defined in two ways, Clausius' entropy of fusion ($\Delta S_f$) based on macroscopic heat transfer ($Q$) at melting temperature $T_m$ (Eq. 1) and Boltzmann's equation for molecular entropy of disorder ($\Delta S_d$) based on the gas contant $R$ and $W_d$ – an increase in the number of microscopic states upon disordering (Eq. 3). Boltzmann's definition is independent of temperature in its direct formulation. While the entropy change provides fundamental structural information in basic research (*1*), applied materials design and processing (*2*), and other fields (*3*), its measurement of nano-scale crystals is a challenge. This is because the Clausius approach requires measuring of $Q$ and $T_m$ on the macroscopic level, while the experimental determination of $W_d$ is not practicable. Our recent statistical mechanical study of the chemical kinetics of electron/molecule interactions (*4*, *5*) led us to speculate that the rate of the decrease of the electron diffraction (ED) signal ($k_I$) (Figs. 1A,B) (*6*), does not merely indicate the rate of the decrease ($k$) of the number ordered molecules in the crystal (*7*, *8*), but also provides a novel method to quantify $W_d$. In our study of a femtogram quantity of ten different crystals (Figs. 1C,D) at 100–296 K under kinematic scattering conditions (*9*), we found that the frequency factor $A$, determined from $k$ over various ambient temperatures ($T$) using the Arrhenius equation (Eq. 3), represents the product of $W_d$ and the cross-section of inelastic electron scattering ($A_{INT}$) (Eq. 4). In essence, $A$ reflects both the molecular degree of freedom and the probability of electron-molecule collisions. From Eq. 4 and the Boltzmann equation (Eq. 2), we obtain Eq. 5 which enabled us to calculate $\Delta S_d$ from $A$ and $A_{INT}$. We report herein that the $\Delta S_d$ values determined for diverse structures match closely with the reported Clausius entropy of fusion ($\Delta S_f$) measured at their respective $T_m$ of >450 K for the ten crystals (Eq. 5, Fig. 1C). An additional nine crystals, including proteins, tRNA, and paraffin shown in Fig. 1D, exhibited similar $W_d$ and $\Delta S_d$ values, attesting to the broad applicability of this new method for probing the degree of freedom of the constituents of crystals. The method thus connects Clausius and Boltzmann experimentally via Eq. 6, and supplements the Clausius approach, being applicable to a femtogram quantity of thermally unstable crystals biomolecular crystals.

Using Eqs. 3-5, we modify the Arrhenius-based disorder rate $k$ (Eq. 3) to derive Eq. 7, showing $k$'s direct link with $W_d$ and the entropy of fusion ($\exp(0.95\Delta S_f/R)$). For flexible molecules like peptides and proteins (see Figs. 2J, 2L), we find negligible activation energy ($E_a \approx 0$), simplifying to Eq. 8, indicating that $k$ is dependent on $W_d$ but independent of $T$. Recent report by Harrowell reported that crystal growth rate from melt ($k_{growth}$) is tied to $\exp(-\Delta S_f/R)$ (Eq. 9, incorporating diffusion coefficient $D$, displacement $\lambda$, and crystal lattice spacing $a$) (*10*). The inverse relation between Eqs. 6 and 8, regarding entropy, reveals that crystal disordering and crystallization are reciprocal processes driven by entropy changes, yet with opposing effects. Consequently, Eqs. 7 and 9 endorse the common experimental observation that small stiff molecules crystallize faster than large and floppy ones, and vice versa for disordering.

(Eq. 1)   $\Delta S_f = Q/T_m$   (Clausius equation)

(Eq. 2)   $\Delta S_d = R\ln(W_d)$   (Boltzmann equation)

(Eq. 3)   $k = A \exp(-E_a/RT)$   (Arrhenius equation)

(Eq. 4)   $A = A_{INT} W_d$,   $A_{INT} = 6.81 \times 10^{-7}\ (e^-)^{-1}\ nm^2$

(Eq. 5)    $A = A_{INT} \exp(\Delta S_d/R)$

(Eq. 6)    $\Delta S_d = a\Delta S_f$, $a = 0.95 \pm 0.1$

(Eq. 7)    $k = A_{INT} W_d \exp(-E_a/RT) = A_{INT} \exp(\Delta S_d/R) \exp(-E_a/RT) = A_{INT} \exp(0.95 \Delta S_f/R) \exp(-E_a/RT)$

(Eq. 8)    $k = A_{INT} W_d = A_{INT} \exp(\Delta S_d/R) = A_{INT} \exp(0.95 \Delta S_f/R)$

(Eq. 9)    $k_{growth} = \dfrac{Da}{\lambda^2} \exp(-\Delta S_f/R)$

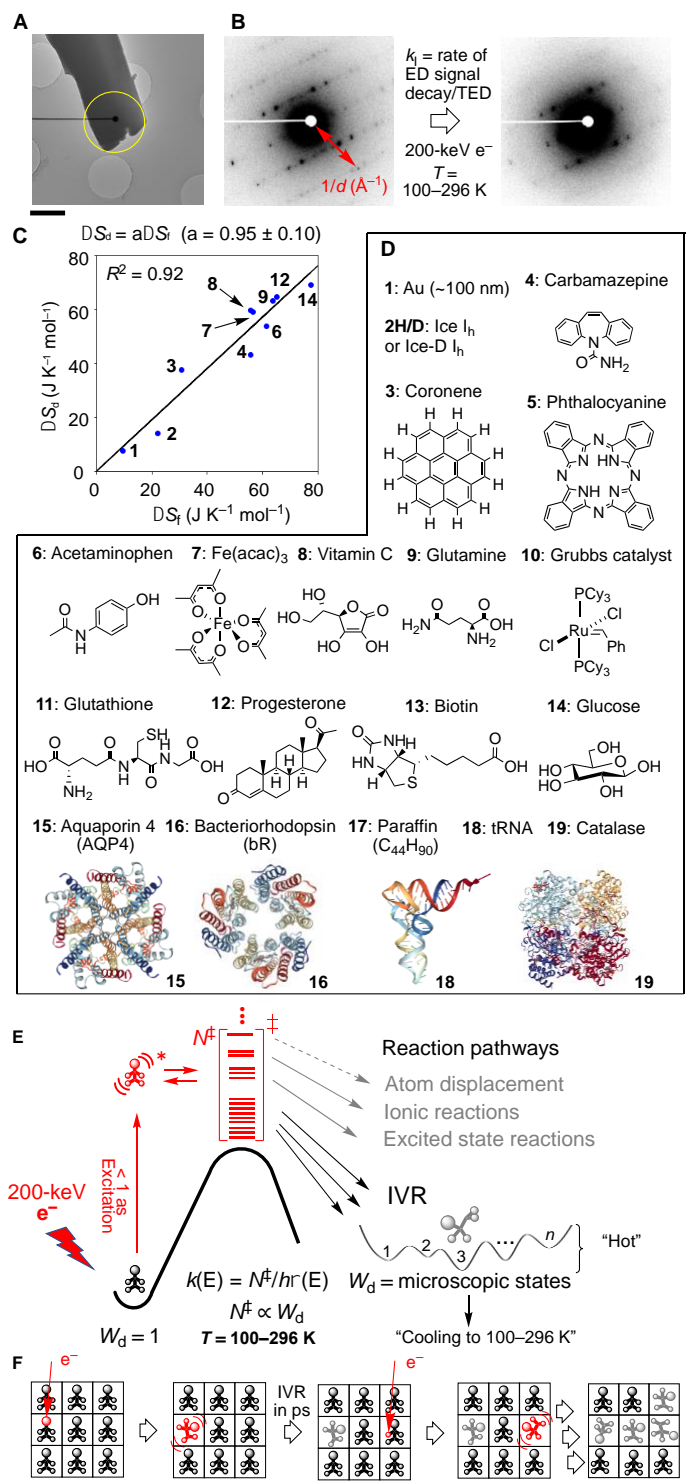

**Fig. 1. Entropy of disorder by ED kinetics for organic, inorganic, and protein crystals.** (**A**) Electron microscopic image of a phthalocyanine (Pc) crystal **5** used for the ED kinetic study. The yellow circle highlights the area selected for ED. Scale bar = 1 μm. (**B**) ED signal decay for the Pc crystal in **A**. TED: total electron dose. (**C**) The $\Delta S_d/a$ data correlated with the molar entropy of fusion $\Delta S_f$ reported for 10 crystals. The sources of the $\Delta S_f$ data are shown in the Supplementary Material (SM), and the literature data intrinsically contain sizable errors as discussed in the caption of Fig. 3. (**D**) Crystal specimens studied. (**E**) RRKM representation of the electron-impact excitation and relaxation of a molecule via diverse reaction pathways, where intramolecular vibrational energy redistribution (IVR) is dominant under low-temperature conditions. $k(E)$ is the RRKM rate constant for the reaction at a given internal energy ($E$) at the energized state. $N^{\ddagger}$ is the number of quantum states at the transition state, leading to the creation of the microscopic states that amount to $W_d$, $h$ is Planck's constant, and $\rho(E)$ is the density of states at $E$. (**F**) Molecule-by-molecule excitation followed by IVR inducing first-order ED signal decay. This illustration depicts only the electrons (in red) impacting molecules, a minor portion of the numerous electrons passing through the crystal.

**Kinetics of ED signal decay**

The decrease in ED intensity of organic crystals with increasing total electron dose (TED = $t$, under constant electron dose rate, EDR) is well known (Fig. 1A,B), but has rarely been the subject of quantitative kinetic studies (*11*, *12*, *13*). The decay of ED signal intensity ($I_t/I_0$) corresponds to the reduction in the number of ordered molecules. In Eq. 10, $I_t$ is proportional to the square of the product of the structure factor ($f$) and the number of unit cells ($N_{UC,t}$) at a TED ($t$) (*7*, *8*), represented as $N_t/Z$, where $N_t$ is the number of ordered molecules in the crystal, and $Z$ is the number of molecules in a unit cell. Thus, the natural logarithm of the ratio ($\ln(I_t/I_0)$) is proportional to $\ln(N_t/N_0)$ (Eq. 11), assuming constant $f$ during the initial stage of electron irradiation, where $t$ is still small. Through experimental observations, we established a first-order relationship between $\ln(I_t/I_0)$ and $t$, enabling the determination of the rate constant ($k_I$) for the signal decrease (Eq. 12; e.g., Fig. 2A and 2H). By applying Eqs. 10 and 11, we derived $k = k_I/2$, with $k$ representing the rate constant for the decrease in the number of ordered molecules ($N_t/N_0$) with respect to $t$ (Eq. 13). Thus, we can obtain $k$ from the $t$-dependent decrease of $\ln(I_t/I_0)$.

(Eq. 10) $\quad\quad\quad\quad I_t \propto (fN_{UC,t})^2 = (fN_t/Z)^2$

(Eq. 11) $\quad\quad\quad\quad \ln(I_t/I_0) = 2\ln(N_{UC,t}/N_{UC,0}) = 2\ln(N_t/N_0)$

(Eq. 12) $\quad\quad\quad\quad \ln(I_t/I_0) = -k_I t$

(Eq. 13) $\quad\quad\quad\quad \ln(N_t/N_0) = -kt = k_I t/2$

Fryer *et al.* in 1992 performed the ED signal analysis for aromatic crystals and reported an extremely large $E_a$ value of 820 kJ mol$^{-1}$ (Fig. 2D, Fig. S1 in the SM) (*14*). At the start of our investigation, we noticed that this value was incorrect because their analysis deviated from the standard kinetics method (*15*), and found that the $E_a$ values for organic crystals are generally 1 kJ mol$^{-1}$ or less (see below). We found that the present experiments performed under kinematic scattering conditions using <1-µm thick organic crystal and high-speed and monochromatic electrons at low EDR conform to the theoretical framework of unimolecular reaction – the statistical kinetics theory of Rice–Ramsperger–Kassel–Marcus (RRKM) theory (*16*, *17*) – a kinetic expression of Boltzmann's equation for understanding how the distribution of energy among the various microscopic states of the molecule influences the reaction rate. In this framework, we consider that each molecule in crystal is excited by an electron and undergoes intramolecular vibrational energy redistribution (IVR) (Figs. 1E, F).

Thus, a 200-keV electron with a short de Broglie wavelength (2.5 picometers) and a speed of 70% of light interacts with one atomic nucleus in a molecule, and transfers its energy to a molecule within less than an attosecond where the molecule can have only one microscopic state ($W_d = 1$, Fig. 1E) (*18*). The whole molecule vibrates on a nanosecond scale before rapid cooling to the ambient temperature $T$ (4–300 K). The experimental observation suggests that the molecule preferentially undergoes vibrational relaxation among other higher energy paths, probably in less than nanoseconds (*19*, *20*). For van der Waals crystals, to which most organic crystals belong, the energy given by the electron to a single molecule does not go beyond the molecular boundary (Fig. 1F). The low EDR guarantees that each event occurs independently of others, resulting in the first order kinetics. The validity of such a hypothesis that one electron energizes only one molecule at a time was proved previously in the kinetic analysis of electron-induced [2 + 2] reaction of van der Waals dimer $C_{60}$ (*5*) and other unimolecular reactions in the gas phase (*21*).

The theory states that the RRKM rate constant of the excited molecule is associated with the number of available quantum states, denoted as $N^\ddagger$, at a given energy ($k(E) = N^\ddagger/h\rho(E)$) (Fig. 1E), where $N^\ddagger$ is proportional to the microscopic states of $W_d$ available to the product. For organic molecules having many C–H and rotatable bonds, the IVR paths are the most populated among all pathes, including IVR, excited state, ionic, and atomic displacement paths available for electron-impact excitation (Fig. 1E). IVR leads to the formation of numerous conformers in crystal. For comparison, $C_{60}$ molecules lacking C–H bonds react predominantly via excited and ionic reaction paths to form [2 + 2]dimers (*4*, *5*). This description of the quantum state of an excited organic molecule allows us to determine $W_d$ from quantitative kinetic analysis of ED signal decay. Then, using the Arrhenius and Boltzmann equations, Eqs. 2 and 3, we can determine $\Delta S_d$ (*22*, *23*, *24*). It is worth noting that light, used in earlier studies of crystal melting, energizes crystal constituents via electric dipole interaction (*25*, *26*, *27*).

We noted several advantages of the statistical mechanical approach using a transmission electron microscope (TEM) over the Clausius calorimetric analysis. The method requires a single crystal of femtogram quantity (Fig. 1A), instead of macroscopic amounts of pure crystals (*28*). It has broader applications to crystals that sublime instead of melt (e.g., **5**) or decompose below $T_m$ (the Grubbs Ru carbene catalyst **10**). We can also determine $W_d$ for various macromolecules such as paraffin (**17**), proteins (**15**, **16**, **19**) and RNA (**18**) (Fig, 1D). The technique accommodates flexible specimen selection: We can work with a mixture of different crystals by choosing target specimens under the microscopic mode of the TEM (Fig. 1A). Furthermore, the measurement can be performed on a common TEM equipped with a variable-temperature probe.

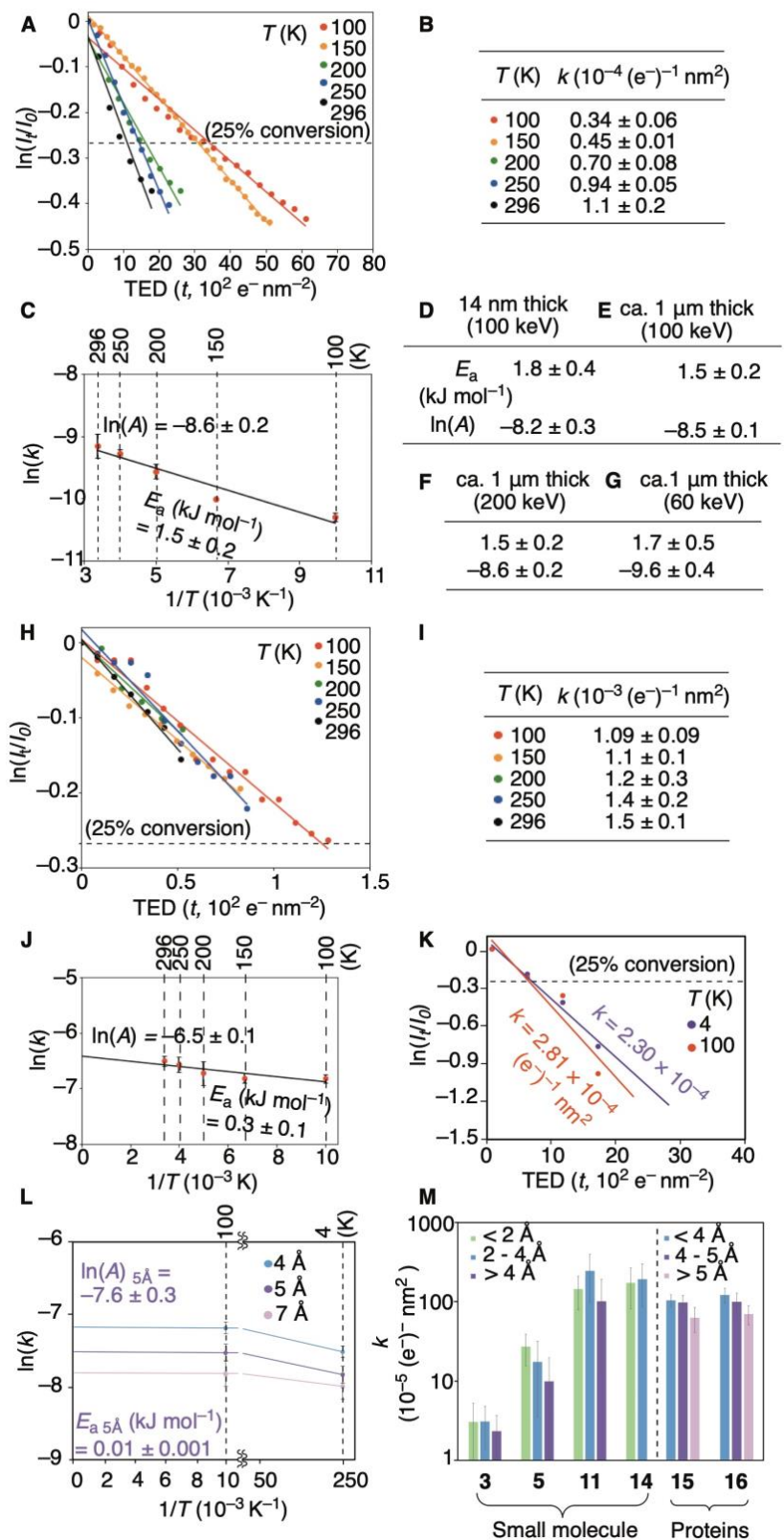

**Fig. 2. Variable-voltage kinetic study of crystal disordering.** (**A**) ED intensity decay for an approximately 1-μm thick Pc crystal at 200 kV. TED = 0 was set as soon as the system stabilized. (**B**) The rate constant ($k$) was obtained via linear regression analysis for a Pc crystal at

200 kV. (**C**) Arrhenius plot for Pc crystals at 200 kV. (**D**) Kinetic data for Pc thin film at 100 kV. (**E**) Kinetic data for Pc crystal at 100 kV. (**F**) Kinetic data for Pc crystal at 200 kV. (**G**) Kinetic study for Pc crystal at 60 kV. Measurement uncertainties listed in **D**–**G** were estimated from the standard deviation. (**H**) ED intensity decay for approximately 1-μm thick glutathione crystal at 200 kV, illustrating the negligible $T$ dependence of $k$. (**I**) The rate constants $k$ were obtained via linear fitting for glutathione crystal **11** at 200 kV. (**J**) Arrhenius plot and kinetic data for **11** at 200 kV. (**K**) ED intensity decay for aquaporin-4 crystal **15** at 300 kV. (**L**) Arrhenius plot and kinetic data for **15** at 300 kV. (**M**) Dependence of the rate constant on the $d$-spacing (interplanar spacing, cf. Fig. 1B) in representative compounds. See the SM for details. The numbers in the x-axis refer to the compounds shown in Fig. 1D. Error bars correspond to standard deviation.

Upon re-analysis of the 1992 Fryer's 820 kJ mol$^{-1}$ $E_a$ data (*14*), we immediately noted that they made the mistake of using the TED up to total loss of ED signal to quantify the progress of the reaction (Fig. 2D, Fig. S1 in the SM). Instead, we performed the standard kinetic analysis focusing on the initial rate (*15, 29*), while essentially following Fryer's experimental protocol. We analyzed approximately 1-μm thick Pc crystals using 60-, 100-, and 200-kV electron beams (Figs. 1A,B). ED patterns were analyzed by irradiating a selected area of approximately 2 μm$^2$, marked by the yellow circle in Fig. 1A (for gold nanoparticles, we used specimens of 100 nm in diameter, cf. Fig. S2). For the Pc crystal, EDR was maintained low at $2.4 \times 10^2$ e$^-$ nm$^{-2}$ s$^{-1}$ (Fig. 1A). EDR of 2.4 to $33 \times 10^2$ e$^-$ nm$^{-2}$ s$^{-1}$ at 200 kV was used in other experiments on crystals **1** to **14**, depending on crystal size and type.

The rate constant, $k$, was calculated across a temperature range of $T$=100–296 K for samples **1** to **14**. The decay of the signal followed first-order kinetics for at least a 20% decrease in signal (refer to Fig. 2A), indicating a consistent structure factor $f$. This consistency was further evidenced by the stable $d$-spacing in the ED signals of phthalocyanine (Pc, **5**), glutamine (**9**), and glutathione (**11**) at 100 K, up to a 20% signal reduction as shown in Fig. 3. However, beyond 50% conversion, the $d$-spacing began to vary, suggesting a change in $f$ and a deviation from first-order kinetics. The same rate constant $k$ was obtained regardless of the crystal's initial orientation, but the rotation of the specimen altered the signal intensities. All crystals examined between 100–296 K showed no evidence of crystal-wide melting, even after the complete disappearance of ED signals. The rate constant, $k$, for the reduction of $N_t/N_0$ was determined over the temperature range of $T = 100–296$ K (Fig. 2B). The rate increased only 3.2 times when the temperature was raised from 100 K to 296 K for Pc. The decrease in $\ln(I_t/I_0)$ yielded identical first-order kinetic parameters, regardless of the crystal's initial orientation.

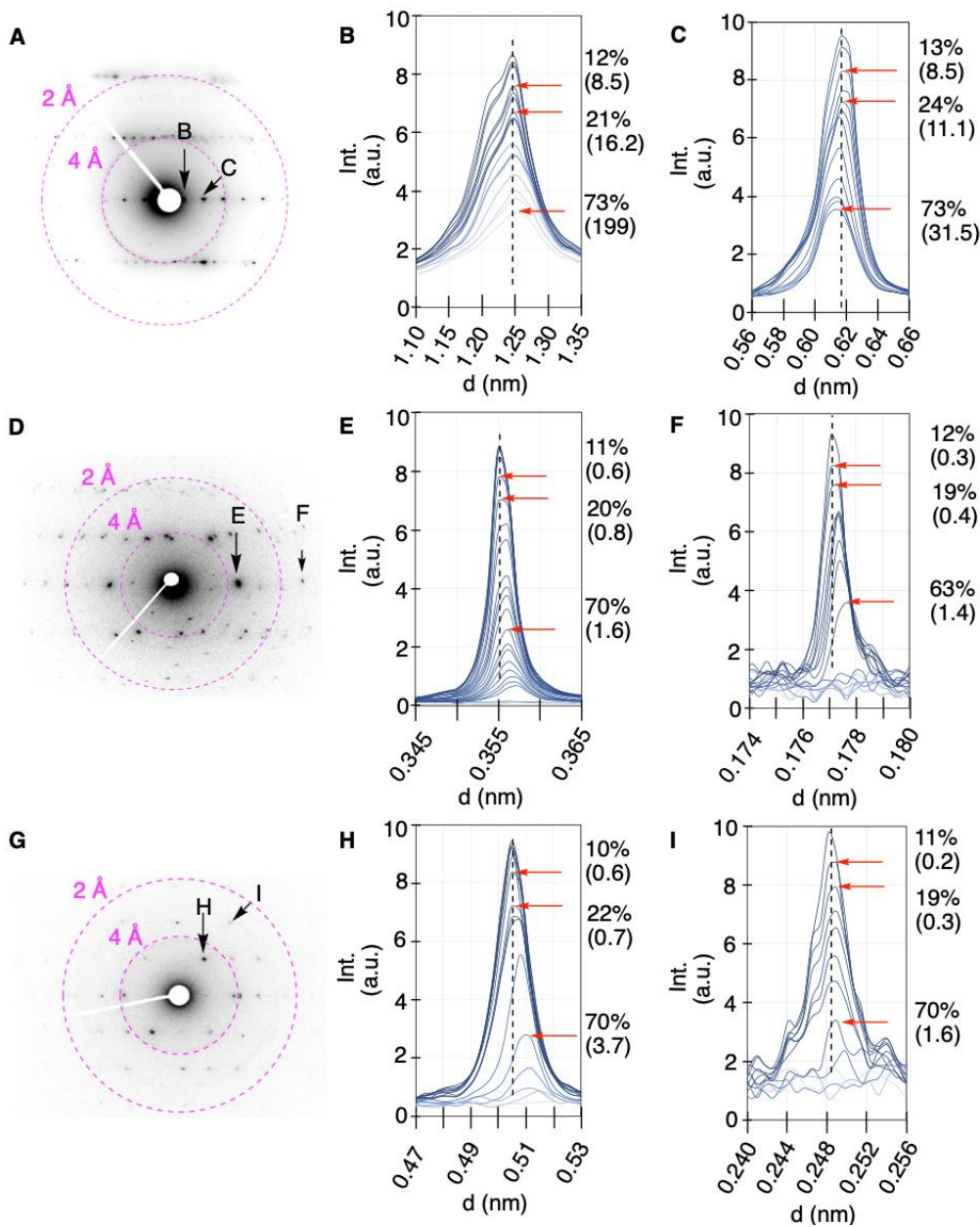

**Fig. 3.** The dependence of *d*-spacing of some ED signals on the % decrease of the ED signal intensity measured for Pc (**5**), glutamine (**9**), and glutathione (**11**) at 100 K. (**A**) The selected diffracted spots of Pc crystal for analysis. (**B**,**C**) The intensity/*d*-spacing correlation analysis for Pc. (**D**) For glutamine **9**. (**E**,**F**) Correlation analysis. (**G**) For glutathione **11**. (**H**,**I**) Correlation analysis. The *d*-spacing remains unchanged approximately up to 20% signal decrease, and then may change as disordering progresses. The data in parenthesis refer to the total electron dose ($t$ in e$^-$ Å$^{-2}$).

The insensitivity of kinetic parameters, $E_a$ and $\ln(A)$, to specimen thickness supports the fulfillment of kinematic ED conditions. Thus, by analyzing Fryer's raw data on the initial rate of $\ln(I_t/I_0)$ for a 14-nm thick crystalline film of Pc **5** at 100 kV, we determined $E_a$ to be $1.8 \pm 0.4$ kJ mol$^{-1}$ (Fig. 2D). This value is in agreement with the 1.5 kJ mol$^{-1}$ data obtained for a 1-μm thick crystal at 100 kV (Fig. 2E). The lack of dependence on thickness indicates that secondary electrons have a negligible impact on the decay of the ED signal. Practically, these findings affirm the method's suitability for nano-scale applications (*30*). It's worth noting that Pc crystal, being unusually resilient among organic crystals and not melting before sublimation, has not had its $\Delta S_f$ reported. $E_a$ remained unchanged (1.5–1.7 kJ mol$^{-1}$) when varying the acceleration voltage from 200 to 60 kV. However, the frequency factor $A$ decreased to about one-third of its value at 100 and 200 kV when reducing the voltage to 60 kV (Fig. 2E, F, G). This implies that the 60-keV electrons supplied energy ($E$ in Fig. 1E) only enough to explore one-third of the quantum states at the transition state ($N^{\ddagger}$), in contrast to the 100- and 200-keV electrons. This suggests in turn that 200-keV (or higher) electrons can provide sufficient energy for the molecule to explore the entire range of microscopic states, thereby yielding reliable $W_d$ data. As a result, we opted for 200-keV electrons in the subsequent study. The ED studies on macromolecules were performed at 300-keV electrons.

In Fig. 2H,I, we show that the signal decay rate $k$ for glutathione **11** under 200 keV electron irradiation was much less depending on the ambient temperature ($T$, 100–296 K) than that for the Pc crystal (Fig. 2B). As a result, the activation energy ($E_a$) of **11** was only 0.3 kJ mol$^{-1}$, as shown by the very low slope in the Arrhenius plot (Fig. 2J). In Figs. 2K-L, we show the rate constant of the decay of 5 Å signals for crystals of a membrane protein, aquaporin-4 **15**, and the Arrhenius plot under 300-keV electron irradiation at 4 K and 100 K. The decay rate $k$ shows even smaller $T$-dependence (Fig. 2K), and consequently, $E_a$ of 0.01 kJ mol$^{-1}$ is nearly zero (Fig. 2L). Thus, the 300-keV electron beam is essentially the only energy source for the disordering of these floppy molecules instead of the environmental temperature ($T$). For comparison, we note that $E_a$ for aliphatic C–C bond rotation is 10–20 kJ mol$^{-1}$.

In Fig. 2M, we compare the disordering rate $k$ at 100 K among compact π-stacked molecules (**3** and **5**), glutathione (**11**), glucose (**14**), and proteins (**15** and **16**). The $k$ values for molecules **11**–**16** are similar to each other, while those of the **3** and **5** were smaller. For the protein crystals, we find smaller $k$ at short $d$-spacing (<4 Å), indicative of larger $W_d$ for peptide side chains, in light of Eq. 8 for the near-zero $E_a$ crystals such as **15** (see above).

**Table 1 Arrhenius and molar entropy parameters determined by ED analysis and reported in the literature. 1–14**: The $E_a$, ln($A$), $W_d$ data for small molecules measured at 100–296 K using 200-keV electrons (Fig. 1C, Figs. S3–22). **15–19:** The data for macromolecules are calculated from the published $t$-dependence of the ED signal decay measured at 4–80 K (AQP4 and bR), 8–300 K (tRNA and catalase), or 13–80 K (paraffin, C$_{44}$H$_{90}$) using 300-keV electrons (*31*, *32*). The $\Delta S_d$ values, derived from the partial structural features of biomacromolecules and their associated $W_d$, are indicated in brackets because the molar entropy cannot be defined for the partial ED data. The molar entropy of fusion data $\Delta S_f$ (blue) are taken from the literature shown in the SM. Sub = sublimation, Dec = decomposition. Unit for $E_a$ = kJ mol$^{-1}$ and molar entropy = J K$^{-1}$ mol$^{-1}$.

| | $E_a$ | $\ln(A)$ | $W_d$ | $\Delta S_d$ | $\Delta S_f$ | $\Delta S_f - \Delta S_d$ |
|---|---|---|---|---|---|---|
| **1**: Gold | 1.8 ± 0.9 | –13.3 ± 0.6 | 2.6 ± 1.6 | 7.5 ± 5.0 | 9.4 | +1.9 |
| **2H**: Ice $I_h$ | 3.2 ± 1.1 | –12.5 ± 1.0 | 6.0 ± 6.3 | 14.1 ± 8.3 | 22.0 | +7.9 |
| **2D**: Ice–D $I_h$ | 3.1 ± 1.5 | –12.4 ± 1.5 | 6.7 ± 10.5 | 15.0 ± 12.5 | 22.2 | +7.2 |
| **3**: Coronene | 1.5 ± 0.1 | –9.66 ± 0.04 | (1.2 ± 0.1) × $10^2$ | 37.7 ± 0.3 | 30.6 | -7.1 |
| **4**: Carbamazepine | 0.9 ± 0.2 | –9.0 ± 0.2 | (2.4 ± 0.5) × $10^2$ | 43.2 ± 1.7 | 55.5 | +12.3 |
| **5**: Phthalocyanine | 1.5 ± 0.2 | –8.6 ± 0.2 | (3.6 ± 0.8) × $10^2$ | 46.5 ± 1.7 | Sub. | – |
| **6**: Acetaminophen | 1.1 ± 0.1 | –7.74 ± 0.08 | (9.0 ± 0.8) × $10^2$ | 53.7 ± 0.7 | 61.3 | +7.6 |
| **7**: Fe(acac)$_3$ | 2.5 ± 0.5 | –7.1 ± 0.4 | (1.8 ± 0.7) × $10^3$ | 59.0 ± 3.3 | 56.6 | –2.4 |
| **8**: Vitamin C | 0.5 ± 0.1 | –7.0 ± 0.1 | (2.0 ± 0.2) × $10^3$ | 59.8 ± 0.8 | 55.7 | –4.1 |
| **9**: Glutamine | 1.4 ± 0.3 | –6.6 ± 0.2 | (3.0 ± 0.6) × $10^3$ | 63.2 ± 1.7 | 63.7 | +0.5 |
| **10**: Grubbs catalyst | 1.6 ± 0.3 | –6.5 ± 0.2 | (3.3 ± 0.7) × $10^3$ | 64.0 ± 1.7 | Dec. | – |
| **11**: Glutathione | 0.3 ± 0.1 | –6.5 ± 0.1 | (3.3 ± 0.3) × $10^3$ | 64.0 ± 0.8 | Dec. | – |
| **12**: Progesterone | 0.7 ± 0.2 | –6.4 ± 0.2 | (3.7 ± 0.8) × $10^3$ | 64.8 ± 1.7 | 65.0 | +0.2 |
| **13**: Biotin | 0.9 ± 0.1 | –6.0 ± 0.1 | (5.6 ± 0.6) × $10^3$ | 68.1 ± 0.8 | Dec. | – |
| **14**: Glucose | 1.2 ± 0.4 | –5.9 ± 0.3 | (6.2 ± 2.0) × $10^3$ | 69.0 ± 2.5 | 77.3 | +8.3 |
| **15-9Å**: AQP4 (9 Å) | –0.004 ± 0.012 | –8.3 ± 0.2 | (5.0 ± 1.0) × $10^2$ | 49.0 ± 1.7 | — | |
| **16-9Å**: bR (9 Å) | 0.016 ± 0.008 | –8.1 ± 0.1 | (6.1 ± 0.6) × $10^2$ | 50.7 ± 0.8 | — | |
| **17**: Paraffin (C$_{44}$H$_{90}$) | 0.056 | –7.9 | 1.6 × $10^3$ | 52.4 | — | |
| **15-5Å**: AQP4 (5 Å) | 0.011 ± 0.004 | –7.6 ± 0.1 | (1.0 ± 0.2) × $10^3$ | 54.8 ± 0.8 | — | |
| **16-5Å**: bR (5 Å) | 0.021 ± 0.007 | –7.5 ± 0.1 | (1.2 ± 0.1) × $10^3$ | 55.7 ± 0.8 | — | |
| **15-4Å**: AQP4 (4 Å) | 0.008 ± 0.008 | –7.4 ± 0.1 | (1.3 ± 0.3) × $10^3$ | 56.5 ± 0.8 | — | |
| **16-4Å**: bR (4 Å) | 0.022 ± 0.008 | –7.3 ± 0.1 | (1.4 ± 0.2) × $10^3$ | 57.3 ± 0.8 | — | |
| **18**: tRNA (9 Å) | 0.02 ± 0.09 | –6.9 ± 0.9 | (2.2 ± 2.0) × $10^3$ | 60.7 ± 7.5 | — | |
| **19**: Catalase (5.5 Å) | 0.15 ± 0.07 | –5.2 ± 0.6 | (1.3 ± 0.8) × $10^4$ | 74.8 ± 5.0 | — | |

## Correlating ln(*A*) with entropy

Having established access to $E_a$ and $\ln(A)$, we measured their values for crystals of the 19 diverse substances in Fig. 1D, and show them in Table 1 together with $W_d$, $\Delta S_d$, $\Delta S_f$, and the difference $\Delta S_f - \Delta S_d$ for 10 of them reported in the literature. The $E_a$ data for organic crystals were nearly zero, 0.008–1.5 kJ mol$^{-1}$, reflecting the temperature-insensitivity of the decay rate (cf. Fig. 2H). The challenge was to decipher the physicochemical significance of $\ln(A)$, which varies widely between –13.3 for gold and –5.9 for glucose. To this end, we compared $\ln(A)$ to the 14 published parameters of the crystals, molecular properties, and electron/molecular interactions summarized in SM (Fig. S23, Table S1), and we observed that only the molar entropy of fusion, $\Delta S_f$, demonstrates a significant correlation with $\ln(A)$ (Fig. 4A). The strong correlation between the frequency factor *A* and the entropy change may not be surprising after all because both concepts relate to the issue of probability, as emphasized by Boltzmann. We found that the $E_a$ data displayed little correlation with the 14 parameters examined (Table 1; Fig. S22).

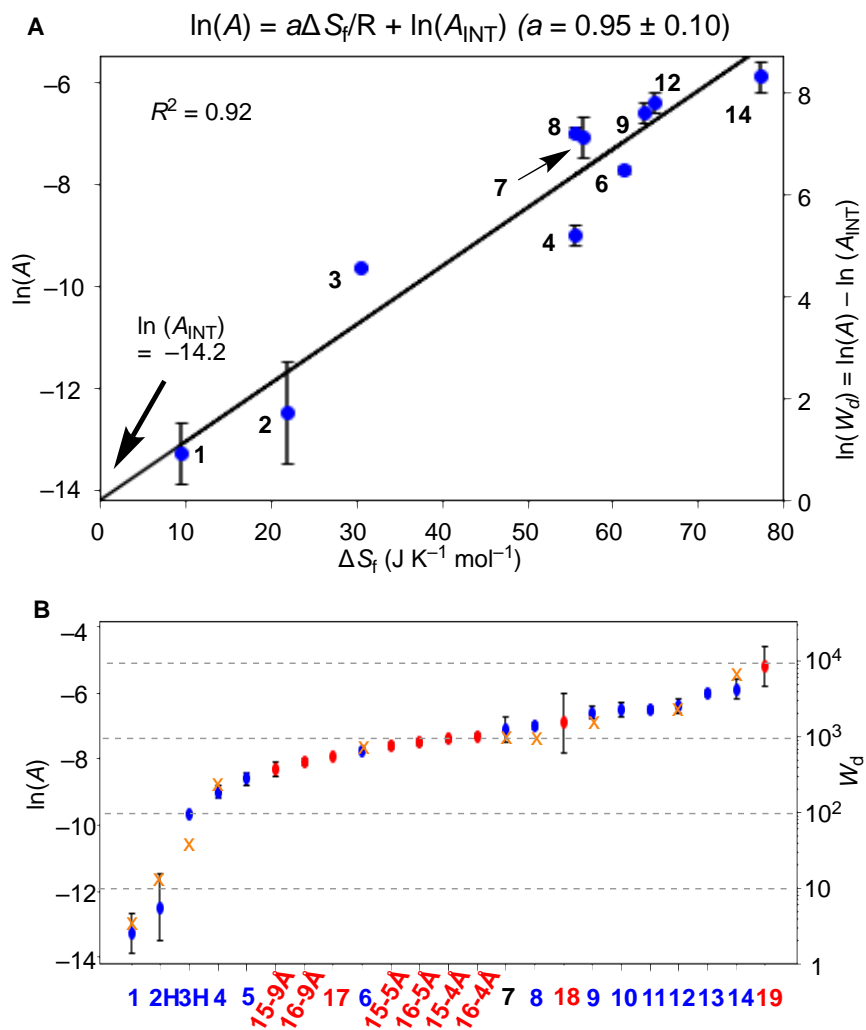

**Fig. 4. Correlation between ln(*A*) and Clausius entropy Δ*S*f and variation of ln(*A*) and *W*d.**
(**A**) ln(*A*)/Δ*S*f correlation. The Δ*S*f data contain a sizable experimental error though not mentioned explicitly in the literature shown in the SM. For example, we selected 30 crystals with one to 10 carbon atoms from reported organic substances (*33*), which were analyzed by multiple melting entropy measurements. The variations in the data ranged from 0.1% to 20.1%, with a standard deviation of 6.6%. The errors for ln(*A*) originate from the Arrhenius plot of the rate ln(*k*) (cf. Fig. 2C). The large errors found for gold (**1**) and ice (**2**) are due to extremely slow signal decay and low conversion. (**B**) ln(*A*) and *W*d plot over the 14 small molecules (blue), and for macromolecules (red) shown in Table 1. Orange cross refers to *W*f calculated from Clausius Δ*S*f and Boltzmann's formulas (Eq. 2).

Fig. 4A and the experimental equation displayed at the top of Fig. 4A present our most significant finding: a statistically significant correlation ($R^2 = 0.92$) between ln(*A*) and the reported Clausius entropy of fusion (Δ*S*f). The slope of the regression line closely approximates the inverse of the gas constant *R* with an additional factor of *a* = 0.95, indicating that the entropy of fusion (Δ*S*f) is, on average, 5% larger than the entropy of disorder (Δ*S*d) as per Eq. 6. It is possibly because entropic contribution of molecular translation is negligible under the low-temperature conditions of our ED study, in contrast to its non-negligible contribution under the Clausius measurements at the melting temperature ($T_m$). The y-intercept of the regression, $\ln(A_{INT})$ represents the dimensions of electron scattering cross-section and the probability of an electron energizing a molecule. Therefore, this experimental equation, reorganized using the Boltzmann equation (Eq. 2), indicating that the frequency factor $A$ is composed of the product of $A_{INT}$ and $W_d$ (Eq. 4). Given the $\ln(A)/\Delta S_f$ correlation in Fig. 4A from which we derived Eqs. 3–5, we reorganize the Arrhenius equation (Eq. 3) to derive Eq. 7. This reformulation indicates a direct relationship among the disorder rate ($k$), $W_d$, and the entropy of disorder ($W_d$ or $\exp(\Delta S_d/R)$), and Eq. 7 tells us that large floppy molecules melt faster than small rigid ones (e.g., Pc).

Figure 4B graphically illustrates the overall trend in the structure – $\ln(A)/W_d$ correlation across various crystals, including gold, small and large organic molecules (depicted in blue), and protein molecules (depicted in red). This visualization provides insight into the degrees of freedom available to the constituents of the crystals and quantifies their propensity for disorder. Notably, the significantly lower ln(A) value for ice, compared to that of floppy organic molecules, suggests that organic molecules, such as proteins embedded within a frozen ice matrix, exhibit a preference for disorder upon exposure to electron irradiation. Table 1 numerically summarizes $W_d$ and molar entropy of disorder $\Delta S_d$ that vary from 2.6 to 6.2 × 10³ and 7.5 to 69.0 J K⁻¹ mol⁻¹, respectively providing insight into the degree of freedom available for the crystal constituents. The top half of Table 1 shows the $\ln(A)$ and $W_d$ data for the 14 specimens, arranged in the order of increasing $\ln(A)$. The difference between the reported entropy of fusion $\Delta S_f$ and $\Delta S_d$ determined by the ED method is shown in the far-right column. The $\Delta S_f$–$\Delta S_d$ gap is small, less than 10%, for floppy organic molecules, rendering the method useful as a practical method for the assessment of melting entropy (Note: The variations in the $\Delta S_f$ data range from 0.1% to 20.1%, with a standard deviation of 6.6%; Fig. 4A caption). The large $\Delta S_f$–$\Delta S_d$ gap for ice suggests a contribution of competing intramolecular relaxation and intermolecular events via hydron bonding network (*34*), and so do the relatively large gaps for crystals of aromatic molecules, **3**, **4**, and **6**, due to strong intermolecular interactions. For gold (*35*), $\Delta S_d$ is 7.5 J K⁻¹ mol⁻¹, similar to the Clausius entropy of $\Delta S_f$ = 9.4 J K⁻¹ mol⁻¹ and Richard's rule value of 8.3 J K⁻¹ mol⁻¹ (*36*).

Table 1 (bottom half) shows the $E_a$ to $\Delta S_d$ data for macromolecules **15** to **19**, derived from ED data at temperatures ranging from 4–300 K and under 300-kV electron irradiation. Remarkably, these macromolecules show similar trends and magnitudes to smaller amide and peptide molecules such as glutathione **11** listed in the table's top half. Specifically, all of the $E_a$ values are near zero, suggesting that the 300-keV electrons effectively disorders the molecules in the crystals, and hence Eq. 8 holds for these crystals. The $\ln(A)$ values fall between –8.3 and –5.2, similar to those of structurally related small molecules. This suggests a common mechanism of disordering across the small and the large molecules, because a highly energetic electron can excite at once all vibrational modes of the molecule – a coupled oscillator, allowing the molecule to explore all conformational possibilities (microstates) during IVR (cf. Fig. 1E). The membrane proteins aquaporin-4 **15** and bacteriorhodopsion **16** showed significant $d$-dependency, reflecting a large difference in $W_d$ for the fragments of shorter (4 Å) and longer (9 Å) periodicity (Fig. 4B).

**Molecular dynamics simulation**

Although the concept of $W$ in the Boltzmann equation is notably abstract, we interpret $W_d$ as the number of unique conformers to be produced when individual molecules, organized within a cold matrix, are subjected to instantaneous superheating through interaction with electrons (Fig. 1F). In accordance with the RRKM framework, $W_d$ is assumed to be proportional to $N^{\ddagger}$ for the superheated molecule (Fig. 1E). To gain qualitative insight into the available number of states for the heated molecule, we conducted molecular dynamics (MD) simulations. These simulations explored the conformational behavior of an energized molecule situated within a cold matrix, utilizing three amide molecules, carbamazepine **4** ($W_d = 2.4 \times 10^2$), acetaminophen **6** ($9.0 \times 10^2$), and glutamine **9** ($3.0 \times 10^3$), whose $W_d$ values span the whole range that we found for small molecules ($1.2 \times 10^2$ to $6.2 \times 10^3$).

We conducted simulations maintaining the crystalline matrix at 300 K (thermal energy of $k_B T = 0.026$ eV, $k_B$ is the Boltzmann constant) and 100 K ($k_B T = 0.0086$ eV) while heating one central molecule to 1000 K ($k_B T = 0.086$ eV) and 2000 K (0.172 eV) (Fig. 5, SM). We observed the dynamics for 0.5 ns, modeling the process depicted in Fig. 1E. To evaluate a qualitative measure of the relative number of microscopic states of each molecule in the crystal, we devised the following procedure: First, we ran simulations of an isolated molecule in the gas phase and characterized its conformational landscape using the molecular dihedral angles as descriptors (see the SM for full simulation details). We then used principal component analysis (PCA) to obtain the linear combination of the descriptors that best separates the different conformers of each molecule (*37*) (Fig. S25–26). We performed simulations of the molecular solid locally heating a single molecule at the center of the crystal to 1000 K or 2000 K while the rest of the environment molecules remained at 300 K or 100 K (Fig. S25). We analyzed the changes in the conformational state of the heated molecule using one principal component (note that we only analyzed intramolecular descriptors).

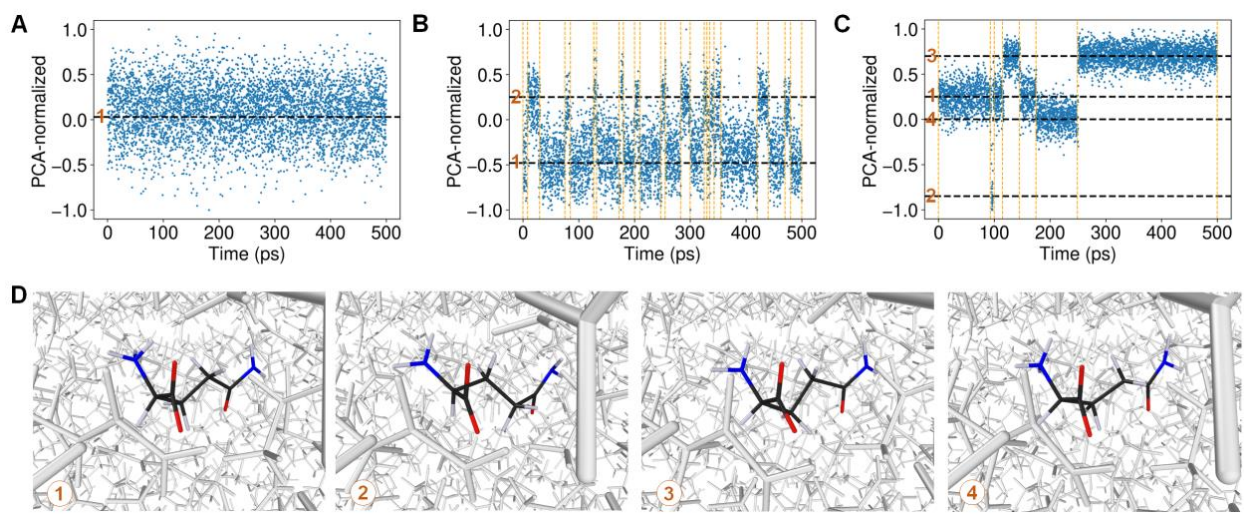

**Fig. 5.** MD simulations of molecular disorder in the crystal. The matrix is kept at 300 K and the molecule at 1000 K. Traces show the time evolution of the first principal component during an MD simulation of 500 ps. The dashed orange lines show transitions. (A) Carbamazepine **4**. (B) Acetaminophen **6**. (C) Glutamine **9**. (D) The time evolution of conformational change of glutamine corresponding to the four conformers marked in panel C. See the SM for computational details (Table S2, Movies S1–S3).

In the simulation, we observed an increase in the number of states in the following order: carbamazepine < acetaminophen < glutamine—in qualitative agreement with the $W_d$ data. During the 0.5 ns simulation, carbamazepine exhibited no transitions; only fluctuations within the same state were observed (Fig. 5A). Acetaminophen showed transitions between two distinct states (Fig. 5B), whereas glutamine showed transitions between three to four distinct states (Fig. 5C). During transitions between the four distinct states, alterations in the molecular conformation of glutamine are evident (Fig. 5D, Movies 1–3 in the SM).

For three different molecules, we obtained the average number of states from 15 independent trajectories, or relative frequency of the appearance of molelcular conformations. The counts were 3.4 ± 0.5 for glutamine, 2 for acetaminophen, and 1 for carbamazepine. Interestingly, these numbers qualitatively align with the ratio of $W_d$ for the three molecules, roughly 12.5:3.75:1 This coincidence suggests a correlation between the molecules' degrees of freedom in crystals and the rate $k$ of molecular disorder shown by Eq. 7. Specifically, more rigid molecules have fewer degrees of freedom and thus disorder slowly. In contrast, more flexible molecules, with greater degrees of freedom, disorder more rapidly, as demonstrated by the correlation between molecular structures and $\ln(A)$ and $W_d$.

One observation is that each primary conformational state contains many short-lived minor states (Fig. 4A and Movies 1–3 in the SM). These are due to the molecule's interactions with the matrix and were not distinctly separated in our PCA analysis (see Fig. S26). It is plausible that these minor states contribute to the large experimental values of $W_d$ and $\Delta S_d$. However, a detailed exploration of the physical significance of $W_d$ falls outside the scope of the current study.

The same trend persists when the environmental molecules are maintained at 100 K (Fig. S27), as well as when the central molecule is heated to 2000 K, while the environmental molecules are maintained at 300 K (Fig. S28). When kept at 300 K, none of the three molecules showed any transitions between states (Fig. S29). Cluster analysis employing the *t*-distributed stochastic neighbor embedding model showed the same qualitative trend found in the PCA (*38*). Using the dihedral angle of the heated molecule as a descriptor, we observed four and two well-distinguished clusters for glutamine **9** and acetaminophen **6**, respectively, whereas carbamazepine **4** showed only one cluster (Fig. S30).

In summary, we have exploited the largely untapped potential of electrons to excite individual molecules vibrationally in a frozen crystal matrix, enabling us to examine their dynamic transitions to an array of microscopic states, that is, conformational possibilities. The credibility of our approach found support in the noteworthy agreement between the determined molar entropy of disorder $\Delta S_d$ and the Clausius entropy fusion $\Delta S_f$ for a variety of organic and inorganic crystals. These observations were interpreted within the framework of the RRKM theory (*19*), and the kinetics based on molecule-by-molecule electronic excitation (*4*, *5*). Given the established validity of RRKM theory in photophysics (*21*), we suggest that this mechanistic framework is equally applicable to the decay of X-ray diffraction signals. Moreover, recent reports on the capabilities of TEM to allow for microscopic analysis of the time evolution of discrete chemical events suggest the feasibility of the statistical mechanics approach to obtaining the kinetic parameters that have conventionally been determined thermodynamically (*39*, *40*, *41*, *42*). This shift is exemplified also in our current study, where upcoming research endeavors will broaden the application of the method and improve the precision of the $\Delta S_d$ data. The reciprocal correlation between electron-induced crystal disorder and crystallization of melt shown by Eqs. 7–9 provides a basis of understanding of crystal nucleation, growth, and disordering (*10*). The data presented herein highlight the importance of the kinetics of diffraction signal decay across a variety of disciplines, including chemical physics, crystallography, structural biology, and nanoscience/technology.